# Optical properties of anatase and rutile TiO$_2$ studied by GGA+U[∀]

Li Jinping[1-3, *], Meng Songhe[1], Qin Liyuan[1], Hantao Lu[2]

*(1 Center for Composite Materials and Structures, Harbin Institute of Technology, Harbin 150080, China)*

*(2 Center for Interdisciplinary Studies & Key Laboratory for Magnetism and Magnetic Materials of the MoE, Lanzhou University, Lanzhou 730000, China)*

*(3 Yukawa Institute for Theoretical Physics, Kyoto University, Kyoto 606-8502, Japan)*

**Abstract:** The optical properties of thermally annealed TiO$_2$ samples depend on their preparation process, and the TiO$_2$ thin films usually exist in the form of anatase or rutile or the mixture of the two phases. The electronic structures and optical properties of anatase and rutile TiO$_2$ are calculated by means of First-principles generalized gradient approximation (GGA) +*U* approach. By Introducing the Coulomb interactions on 3*d* orbitals of Ti atom *($U^d$)* and 2*p* orbitals of O atom *($U^p$)*, we can reproduce the experimental values of the band gap. The optical properties of anatase and rutile TiO$_2$ are obtained by means of GGA+*U* method, well agreeing with experimental results and other theoretical data. Further we present the comparison of the electronic structure, birefringence and anisotropy between the two phases of TiO$_2$.

## 1. Introduction

TiO$_2$, as one of the wide band-gap semiconductors, with stable, nontoxic and efficient photocatalytic activity, has been widely used in photocatalytic, solar battery, sensor, self-cleaning materials and so on [1-4]. TiO$_2$ thin films usually exist in two polymorphs after high-temperature annealing, i.e., the anatase and the rutile, shorted as a- and r-TiO$_2$, respectively [5-9].

In experiments, optical properties of thermally annealed TiO$_2$ have shown intrinsic birefringence and anisortropy in some samples [10]. For example, the fundamental optical properties of optically anisotropic materials such as r-TiO$_2$ have been measured at room temperatures with linearly polarized light of wavelength longer than 110nm [7]. Polarized reflection spectra of single-crystal a-TiO$_2$ have been obtained in a photon energy range from 2 to 25 eV using synchrotron orbital radiations, and the optical properties show anisotropy [8]. Spectroscopic ellipsometry measurements were made on thin-film and single-crystal a-TiO$_2$ using a two-modulator generalized ellipsometer. The results show that the complex refractive indices and dielectric function in a- and r-TiO$_2$ are quite different. Below the band edge, compared with a-TiO$_2$, r-TiO$_2$ has a higher refractive index, as well as greater value of birefringence [9].

Theoretically, the crystal structure, the band structure, and the density of states of a-TiO$_2$ have been analyzed by using first-principles local density approximation (LDA) approach [11]. The resulting band-gap value is 2.25 eV. Near the absorption edge, it shows a significant optical anisotropy between the components with directions parallel and perpendicular to the *c* axis.

In the case of r-TiO$_2$, the structure and electronic properties have been calculated by using "soft-core" ab-initio pseudopotentials constructed within the LDA, and the resulting band-gap value is 2.0 eV. The obtained dielectric function and reflectivity of r-TiO$_2$ for polarization vector also display difference between the directions parallel and perpendicular to the *c* axis [12].

The self-consistent orthogonalized linear-combination-of-atomic-orbitals method in the local-density approximation has been used to study the electronic structure and optical properties of the three phases of TiO$_2$ [13]. The obtained band-gap values of a-TiO$_2$ and r-TiO$_2$ are 2.04 and 1.78 eV, smaller than the experimental data, 3.2 and 3.0 eV, respectively. The theoretical calculations show a little difference in the optical properties of the three phases in TiO$_2$.

Regardless of the above computational efforts, the physical mechanism for the intrinsic birefringence and anisotropy of the optical properties of TiO$_2$ found in some samples remains to be understood. Note that the band gaps obtained by the previous methods are significant lower than the experimental values. So, it is important to investigate the electronic structures of a-/ r-TiO$_2$ and clarify the difference of optical properties based on the first-principles band structure calculations with electronic correlations to be taken into account effectively.

The LDA+*U* method has been employed to investigate the properties of r-TiO$_2$. The results are dramatically improved when additional correlation corrections are introduced on the O 2*p* orbitals in the LDA+$U^d$+$U^p$ approach [14]. The experimental band-gap value (3.20 eV) can be reproduced when $U^d$=8.0 eV and $U^p$=7.0 eV. In a similar approach, the influence of oxygen defects upon the electronic properties of Nb-doped TiO$_2$ was studied, and the effective *U* parameter $U_{eff}$ = 7.2 eV has been used to correct the strong Coulomb interaction between 3d electrons localized on Ti in anatase models [15]. The calculated results show that the anatase NbTi cells are degenerated semiconductors with a typical n-type degenerated characteristic in their electronic structure, which is in good agreement with the experimental evidence that anatase Nb:TiO$_2$ film is an intrinsic transparentmetal. Other researchers set up a model for Zn1-xAgxO (x = 0, 0.0278, 0.0417) to calculate the geometric structure and energy via the method of generalized gradient approximation (GGA+U), showing that the absorption spectrum in these systems all coincide with experimental data[16].

In this paper, we use the GGA+$U$ scheme formulated by Loschen *et al.* [17], to calculate the electric structures and optical properties of r-TiO$_2$ and a-TiO$_2$. The on-site Coulomb interactions of *3d* orbitals on Ti atom ($U^d$) and of *2p* orbitals on O atom ($U^p$) are determined so as to reproduce the experimental value of band gap for the two phases of TiO$_2$. The comparison of the electronic structure, birefringence and anisotropy between the two phases of TiO$_2$ is also presented.

## 2. Computational methodology

Density functional theory calculations are performed with plane-wave ultrasoft pseudopotential, by using the GGA with Perdew-Burke-Ernzerhof (PBE) functional and the GGA+$U$ approach as implemented in the CASTEP code (Cambridge Sequential Total Energy Package) [18]. The ionic cores are represented by ultrasoft pseudopotentials for Ti and O atoms. For Ti atom, the configuration is [Ar] $3d^24s^2$, where the $3s^2$, $3p^6$, $3d^2$ and $4s^2$ electrons are explicitly treated as valence electrons. For O atom, the configuration is [He] $2s^22p^4$, where $2s^2$ and $2p^4$ electrons are explicitly treated as valence electrons. The plane-wave cut off energy is 380eV and the Brillouin-zone integration is performed over the 24×24×24 grid sizes using the Monkorst-Pack method for monoclinic structure optimization. This set of parameters assure the total energy convergence of 5.0×10$^{-6}$ eV/atom, the maximum force of 0.01 eV/A°, the maximum stress of 0.02 GPa and the maximum displacement of 5.0×10$^{-4}$ A°. We calculate the electronic structures and optical properties of a-TiO$_2$ and r-TiO$_2$ by means of GGA without $U$ and GGA+$U^d$+$U^p$ after having optimized the geometry structure. The details of the calculation have been shown elsewhere [19].

## 3. Results and discussion

The space group of a-TiO$_2$ is I4$_1$/amd and the local symmetry is C4h-19, and r-TiO$_2$ is P4$_2$/mnm and D4h-14. The lattice constants *a* and *c* are experimentally determined to be *a*=0.3785 nm and *c*=0.9515 nm, *a*=0.4593 nm and *c*=0.2959 nm [20-21], respectively. The GGA calculation of the perfect bulk a-TiO$_2$ and r-TiO$_2$ is performed to determine the optimized parameters in order to check the applicability and accuracy of the ultrasoft pseudopotential. The optimized parameters are $a$=0.3795 nm and $c$=0.983 7nm, and $a$=0.4645 nm and $c$=0.2968 nm for a-TiO$_2$ and r-TiO$_2$, respectively, in good agreement with experimental and other theoretical values [10-13]. However, the value of the band gap $E_g$ in a-TiO$_2$ and r-TiO$_2$ is around 2.16 eV and 1.85 eV, respectively, smaller than the experimental value of 3.23 eV [22] and 3.0 eV [23]. This is due to the fact that the DFT results often undervalue the energy of 3d orbitals of Ti atom, lowering the bottom level of conduction bands. As a result, $E_g$ of TiO$_2$ obtained by GGA is lower than the experimental one.

In order to reproduce the band gap, we first introduce $U^d$ for *3d* orbitals of Ti atom. Using the experimental lattice parameters, we optimize geometry structure and calculate the band structure and density of state (DOS) of r-TiO$_2$. The band gap $E_g$ obtained from the band structure is shown in Fig. 1(a) as a function of $U^d$. It can be seen that $E_g$ firstly increases, and then drops with increasing $U^d$, showing a maximum value (2.46 eV) at $U^d$ =6.75 eV, where the lattice parameters of the optimized structure are $a$=0.4664 nm and $c$=0.3086 nm. The maximum value is smaller than experimental one (3.0 eV). The saturation of $E_g$ with $U^d$ may be related to the approach of 3d states toward 4s and 3p states, though microscopic mechanism is not yet fully understood. Next, we introduce $U^p$ for *2p* orbital of O atom, while keeping $U^d$ =6.75 eV. The results in Fig. 1(b) shows that $E_g$ monotonically increases with $U^p$. When $U^d$ =6.75 eV and $U^p$ =3.5 eV, the calculated band gap of r-TiO$_2$ is 3.0 eV, well consistent with experiment one, where the lattice parameters of the optimized structure are $a$=0.4664 nm and $c$=0.3082 nm. For a-TiO$_2$, When $U^d$ =7.75 eV and $U^p$ =1.0 eV, the calculated band gap of a-TiO$_2$ is 3.229 eV, well coinciding with experiment one, where the lattice parameters of the optimized structure are $a$=0.38891 nm and $c$=0.9870 nm.

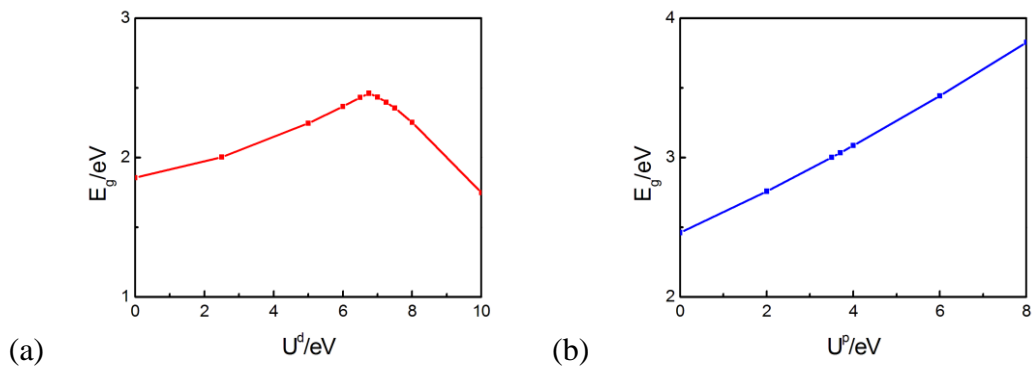

Fig.1 Calculated band gap E$_g$ of r-TiO$_2$ as a function of (a) $U^d$ and (b) $U^p$

By adopting these $U$ values, $U^d$=6.75 eV and $U^p$=3.5 eV, ($U^d$=7.75 eV and $U^p$=1.0 eV for a-TiO$_2$), we perform the GGA+$U$ calculation for r-TiO$_2$. The band dispersion is shown in Fig. 2(a). The bottom of the conduction band is located at the G/B point. Since the bottom shift to higher energy with $U^d$ accompanied by the reconstruction of the conduction band, the separated DOS at 3.03 eV and 6.26 eV obtained by GGA without $U$ (not shown) merge to one sharp structure at 3.74 eV in Fig. 2(b). The conduction band is predominantly constructed by Ti 3d states, while the valence band is by O 2p states as shown in Fig. 2(c) and Fig. 2(d). Therefore, the excitations across the gap are mainly from the O 2p states to the Ti 3d states.

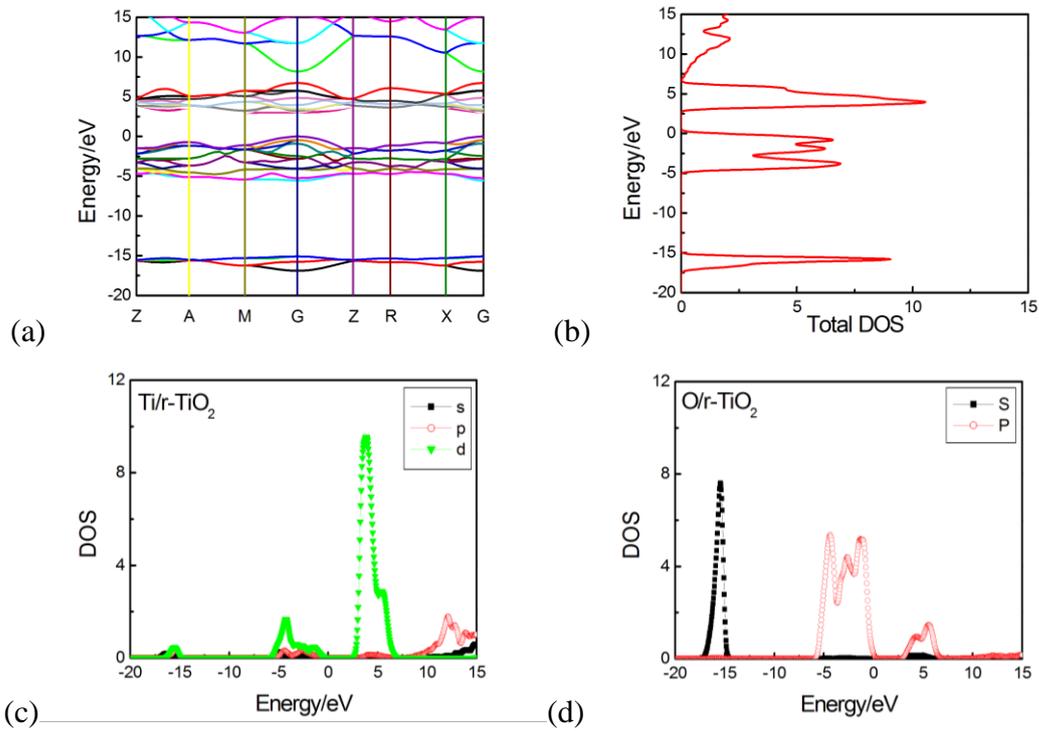

Fig.2 The band structure and DOS of r-TiO$_2$ obtained by GGA+$U^d$+$U^p$ ($U^d$=6.75 eV, $U^p$=3.5 eV). (a) Band structure. The total DOS, the partial DOS of Ti and O atoms are shown in (b), (c) and (d), respectively

Figure 3 shows the comparison of the total DOS between a- and r-TiO$_2$ obtained by GGA and GGA+$U^d$+$U^p$ (($U^d$=6.75 eV, $U^p$=3.5 eV, and $U^d$=7.75 eV, $U^p$=1.0 eV for r-TiO$_2$ and a-TiO$_2$, respectively). From the wide region (shown in Fig. 3(a)), the total DOS of r- and a-TiO$_2$ obtained by GGA and GGA+$U^d$+$U^p$ look quite similar; while zooming into the region around the Fermi surface (Fig. 3(b)), we can see that compared with the GGA-without-$U$ results, a significant change after including $U$ is that the conduction bands are pushed to higher energies, which accordingly produces larger band gaps.

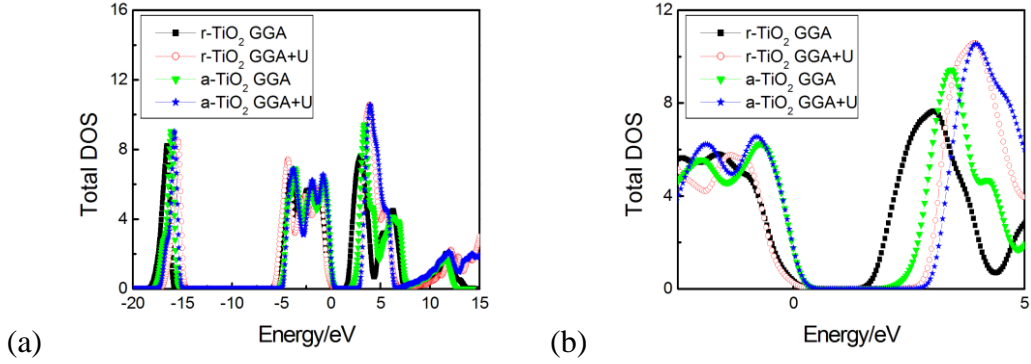

Fig.3 Total DOS of the two phases of $TiO_2$ by GGA without $U$, and GGA+$U^d$+$U^p$ ($U^d$=6.75 eV, $U^p$=3.5 eV, and $U^d$=7.75 eV and $U^p$=1.0 eV for r-$TiO_2$ and a-$TiO_2$, respectively) in a wide region (a), and near Fermi surface (b).

Figure 4 shows the dielectric function of r-$TiO_2$ obtained by GGA +$U^d$ +$U^p$ ($U^d$=6.75 eV and $U^p$=3.5 eV) (a) parallel and (b) perpendicular component. The parallel and perpendicular components of the real part $\varepsilon_1$ have maximums of 12.53 at 3.86 eV and 10.98 at 3.94 eV, respectively. The calculated static dielectric constants are 6.298 and 5.485 along the different directions, coinciding with the experimental value 6.33 [24]. The parallel and perpendicular components of the imaginary part $\varepsilon_2$ show the maximums of 10.28 at 6.74 eV and 9.50 at 4.90 eV, respectively. Other optical properties can be computed from the complex dielectric function [25]. For example, we can obtain the refractive coefficient of r-$TiO_2$, whose parallel and perpendicular component are $n_e$=2.75 and $n_o$=2.46. It is obvious that the values by GGA +$U^d$ +$U^p$ are closer to the experimental values $n_e$=2.7 and $n_o$= 2.4 [7], or $n_e$=2.76 and $n_o$=2.44 [26].

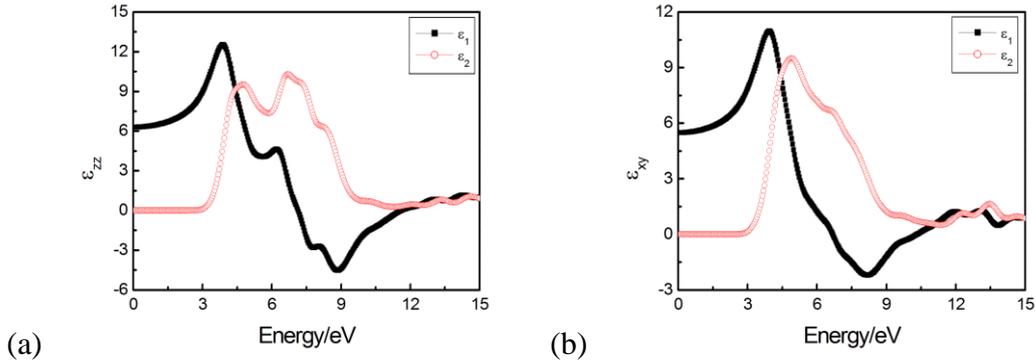

Fig.4 Dielectric function of r-$TiO_2$ obtained by GGA+$U^d$+$U^p$ ($U^d$=6.75 eV and $U^p$=3.5 eV). (a) Parallel and (b) perpendicular component.

Figure 5 shows the dielectric function of parallel and perpendicular component in a-TiO$_2$ obtained by GGA $+U^d +U^p$ ($U^d$=7.5 eV and $U^p$=1.0 eV). From Fig.5, it can be seen the maxmums of the parallel and perpendicular components of the real part $\varepsilon_1$ are 11.79 at 3.75 eV and 12.56 at 3.75 eV, respectively. The static dielectric constants of the two components are 6.137 and 5.995, coinciding with the experimental value 5.62 [24]. The maximums of the parallel and perpendicular components of the imaginary part $\varepsilon_2$ are 10.29 at 4.74 eV and 11.22 at 4.52 eV, respectively. Other optical properties can be computed from the complex dielectric function [25]. The parallel and perpendicular components of the refractive coefficient of r-TiO$_2$ are $n_e$=2.34 and $n_o$=2.29. It is obvious that the values by GGA $+U^d +U^p$ are closer to the experimental values of $n_e$=2.32 and $n_o$=2.28 [27].

From Fig.4 and Fig.5, we can notice the optical anisotropies in both a- and r-TiO$_2$ between parallel and perpendicular components. The rutile form of TiO$_2$ has comparatively high anisotropy, both above and below the band gap, which makes r-TiO$_2$ a very useful optical material [9]. Below the band gap, r-TiO$_2$ has a large birefringence (difference between refractive indexes of the parallel and perpendicular components) $\Delta n = n_e-n_o$=0.29 in our calculation, more agreeable with the experiment data ($\Delta n$=0.27[28] or 0.30 [7]).

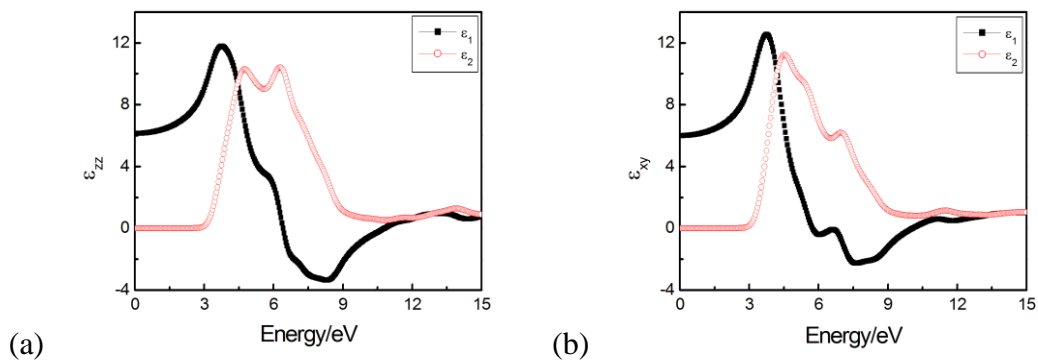

Fig.5 Dielectric function of a-TiO$_2$ obtained by GGA+$U^d$+$U^p$ ($U^d$=7.5 eV and $U^p$=1.0 eV). (a) Parallel and (b) perpendicular component

## 4. Conclusion

The electronic structures and optical properties of anatase and rutile TiO$_2$ are calculated by means of First-principles generalized gradient approximation GGA and GGA) +$U$ approaches. By GGA, the resulting band gaps $E_g$ in a-TiO$_2$ and r-TiO$_2$ are around 2.16 eV and 1.85 eV, smaller than the experimental values. Introducing the Coulomb interactions of 3d orbitals on Ti atom ($U^d$) and of 2p orbitals on O atom ($U^p$), we can reproduce the experimental values of the band gap for a- and r-TiO$_2$. The best values for $U^d$ and $U^p$ are $U^d$=6.75 eV and $U^p$=3.5 eV, $U^d$=7.5 eV and $U^p$=1.0 eV for r- and a-TiO$_2$, respectively. The complex dielectric functions and refractive index of r- and a-TiO$_2$ are calculated. The results both show optical anisotropy in a- and r-TiO$_2$. The r-TiO$_2$ has relatively large anisotropy with birefringence $\Delta n$=0.29, which makes it a useful optical material.